\documentstyle[12pt,epsf]{article}
\input epsf.tex
\catcode`\@=11

\begin{document}
\begin{titlepage}
\samepage{ \setcounter{page}{1} \rightline{OSU-HEP-99-05} \vfill
\begin{center}
{\Large \bf Gauge Coupling Unification\\  with Large Extra Dimensions\\}
\vspace{.3in}
 {\large
Daniel Dumitru\footnote{Email address: ddumitru@uchicago.edu} \normalsize
and \large Satyanarayan Nandi\footnote{Email address: shaown@okstate.edu}
\\}
\vspace{.25in}
{\it Department of Physics, Oklahoma State University,
     Stillwater, OK 74078.\\}
\end{center}
\vfill
\vfill
\begin{abstract}
We make a detailed study of the unification of gauge couplings
in the MSSM with large extra dimensions. We find some scenarios where
unification can be achieved [with $\alpha_3\left(M_Z\right)$ within
$1\sigma$ of the experimental value] with both the compactification scale
and the SUSY breaking scale in the few TeV range. No enlargement
of the gauge group or particle content is needed. One particularly
interesting scenario is when the SUSY breaking scale is larger than
the compactification scale, but both are small enough to be probed
at the CERN LHC. Unification in two scale compactification scenarios is also
investigated and found to give results within the LHC reach.
\end{abstract}
\vfill}
\end{titlepage}


\catcode`@=11
\long\def\@caption#1[#2]#3{\par\addcontentsline{\csname
  ext@#1\endcsname}{#1}{\protect\numberline{\csname
  the#1\endcsname}{\ignorespaces #2}}\begingroup
    \small
    \@parboxrestore
    \@makecaption{\csname fnum@#1\endcsname}{\ignorespaces #3}\par
  \endgroup}
\catcode`@=12


\newcommand{ \slashchar }[1]{\setbox0=\hbox{$#1$}   
   \dimen0=\wd0                                     
   \setbox1=\hbox{/} \dimen1=\wd1                   
   \ifdim\dimen0>\dimen1                            
      \rlap{\hbox to \dimen0{\hfil/\hfil}}          
      #1                                            
   \else                                            
      \rlap{\hbox to \dimen1{\hfil$#1$\hfil}}       
      /                                             
   \fi}                                             %


\newcommand{\hmu}{{\hat\mu}}
\newcommand{\hnu}{{\hat\nu}}
\newcommand{\hrho}{{\hat\rho}}
\newcommand{\hh}{{\hat{h}}}
\newcommand{\hg}{{\hat{g}}}
\newcommand{\hk}{{\hat\kappa}}
\newcommand{\tA}{{\widetilde{A}}}
\newcommand{\tP}{{\widetilde{P}}}
\newcommand{\tF}{{\widetilde{F}}}
\newcommand{\th}{{\widetilde{h}}}
\newcommand{\tp}{{\widetilde\phi}}
\newcommand{\tchi}{{\widetilde\chi}}
\newcommand{\te}{{\widetilde\eta}}
\newcommand{\vn}{{\vec{n}}}
\newcommand{\vm}{{\vec{m}}}
\newcommand{\gsim}{\lower.7ex\hbox{$\;\stackrel{\textstyle>}{\sim}\;$}}
\newcommand{\lsim}{\lower.7ex\hbox{$\;\stackrel{\textstyle<}{\sim}\;$}}


\section{Introduction}
Extra compact dimensions beyond our usual four dimensional
space time appear na\-tu\-rally in string theory. The sizes of
these extra dimensions are not generally fixed by the string
dynamics. These may be close to the inverse of the Plank scale in
which case they will have very little direct phenomenological
implications. However, recent developments in string theory allow
the possibility that the size of these extra dimensions may be very
large \cite{witten,lykken}, such as the inverse of a TeV
\cite{lykken,list1}, or even in the submillimeter range
\cite{dvali}. This has generated the exciting possibility for
their direct phenomenological implications, such as the
modification of Newton's law of gravity in the submillimeter
range \cite{dvali}, the effects in low-energy astrophysical phenomena
\cite{list3}, and in high energy collider physics
\cite{dvali,list4}. Some of the standard model (SM) gauge and
Higgs bosons and their supersymmetric (SUSY) partners may live in
a D brane containing some of these few TeV$^{-1}$ compact
dimensions. Then, the effect of their low-lying Kaluza-Klein (KK)
excitations should be observed in the forthcoming high-energy
colliders either through the direct production of some of these KK
states or through their indirect off-shell effects.

About a year ago, it was pointed out that if the SM particles
propagates into these extra dimensions, then the contribution of
their KK excitations gives additional contributions to the $\beta$ 
functions above the compactification scale, $\mu_0$. This modifies
the running of the gauge couplings from the usual logarithmic
running to an approximate power-law running \cite{dienes}.
Depending on the choice of $\mu_0$, this can lead to the
unification of gauge couplings at a scale much smaller than the
usual grand unified theory (GUT) scale. Typically, the unification occurs at a scale of
$\approx 1.5\mu_0$ to $\approx 20\mu_0$ depending on the number of
extra dimensions, and regardless of the number of fermion families
contributing. This gives the possibility of having the unification
scale as low as few TeV, depending on the choice of $\mu_0$. This
is very exciting, because it not only eliminates the usual gauge
hierarchy problem but also allows the prospect of observing GUT
physics at the forthcoming colliders, such as the CERN Large Hadron Collider (LHC). However, more
detailed study (including the two-loop contributions below
$\mu_0$) shows that such an unification does not occur
\cite{ross}. Using the accurately measured values of
$\alpha_1\left(M_Z\right)$ and $\alpha_2\left(M_Z\right)$ to
determine the unification scale, one finds the values of
$\alpha_3\left(M_Z\right)$ much higher than the experimentally
measured range \cite{ross}, unless the scale of compactification
is very high, such as $10^{12}\:$GeV. (However, it should be noted
that there are several threshold effects which are not included in
the analysis of \cite{ross}. For example, according to
\cite{dienes1}, a 6\% threshold effect at the unification scale
could remedy this situation). Subsequent investigation showed that
the unification with low scale $\mu_0$ can be achieved if one
alters \cite{list5} the minimal supersymmetric standard model (MSSM) spectrum in the extended
$(4+\delta)$-dimensional space, or extend the gauge group with an
intermediate scale \cite{moh}. In theories with extra dimensions,
the effect of higher dimensional operators (induced by the quantum
gravitational effects) on the gauge coupling unification as well
as the possibility of TeV scale unification have also been
investigated \cite{hill}.

In all of these works, it was implicitly assumed that the
supersymmetry is exact at the higher dimensional theory, and it breaks
after the compactification to the four dimensions. Thus the
compactification scale $\mu_0$ was always taken to be higher than the
SUSY breaking scale $\mu_{SUSY}$.

The object of this work is to make a detailed study of the gauge coupling
unification within MSSM with large extra dimensions. Our analysis include
several scenarios not previously considered (but allowed by string
theory). We do not extend the particle content (other than those required
by the extra dimensions) or the gauge group. In addition to the case
$\mu_{SUSY}<\mu_0$, our investigation includes the scenario in which the SUSY
is broken at the higher dimension (before compactification), so that the
SUSY breaking scale is larger than the compactification scale. We are
particularly interested in the cases in which both the compactification
scale as well as the SUSY breaking scale are in the few or few tens of a
TeV scale. We find that for this scenario, ($\mu_0<\mu_{SUSY}$), the
unification of the gauge couplings can be achieved with
$\alpha_3\left(M_Z\right)$ lying within $1\sigma$ of the experimentally
measured range and with both $\mu_0$ and $\mu_{SUSY}$ in the few TeV
scale. Such a scenario can be tested at the LHC. We also study the
unification for the cases where only the gluons or the W, Z, H and/or the
matter contribute above $\mu_0$ and find that unification does not take
place in these cases. Finally, we analyze the scenario in which there are
two scales of compactification, $\mu_{10}$ and $\mu_{20}$. Here we find
two cases which give rise to unification with both $\mu_{SUSY}$ and
$\mu_{10}$ in the few TeV range. Our paper is organized as follows. In
Section \ref{sec2} we discuss the formalism, the relevant equations and
how we do our analysis. In Section \ref{sec3} we consider the case of a
single compactification scale with $\mu_{SUSY}<\mu_0$. Here we compare our
results with those obtained in \cite{ross}. Section \ref{sec4} contains
our most interesting results. Here we give the results for the case
$\mu_{SUSY}>\mu_0$. In Section \ref{sec5} we discuss the results for the
various cases with two scale compactification. Section \ref{sec6} contains
our conclusions.

\section{The formalism}
\label{sec2}
In this section we write down the relevant equations and present the details
of how we perform our calculations leading to the results discussed in Secs. 
\ref{sec3}, \ref{sec4}, \ref{sec5}. The running of the gauge couplings,
$\alpha_i$, up to two loops, is given by

\begin{equation}
\label{e1}
\mu{{d\alpha_i\left(\mu\right)}\over{d\mu}}={{b_i}\over{\left(2\pi\right)}}
\:\alpha_i^2\left(\mu\right)+\sum_{j=1}^3{{b_{ij}}\over{\left(8\pi^2\right)}}
\:\alpha_i^2\left(\mu\right)\alpha_j\left(\mu\right)
\end{equation}
where $b_i$ and $b_{ij}$'s are the one- and two-loop $\beta$-function
coefficients. Eq. \ref{e1} can be integrated iteratively by using the
one-loop approximation for the $\alpha_j$'s in the second term:
\begin{equation}
\alpha_j^{-1}\left(\mu\right)=\alpha_j^{-1}\left(\mu^{\prime}\right)
-{{b_j}\over{\left(2\pi\right)}}\:
\mbox{ln}{{\mu}\over{\mu^{\prime}}}\:\:
.
\end{equation}
The resulting equations give the couplings at a higher scale
$\mu_2$ in terms of the couplings at a lower scale $\mu_1\leq\mu_2$ :
\begin{equation}
\label{e3}
\alpha_i^{-1}\left(\mu_2\right)= \alpha_i^{-1}\left(\mu_1\right)-
{{b_i}\over{\left(2\pi\right)}}\:\mbox{ln}{{\mu_2}\over{\mu_1}}+{1\over{\left(4\pi\right)}}\sum_{j=1}^3
{{b_{ij}}\over{b_j}}\:\mbox{ln}\left({1-{{b_j}\over{\left(2\pi\right)}}\:\alpha_j\left(\mu\right)\mbox{ln}{{\mu_2}\over{\mu_1}}}\right)\: .
\end{equation}
Using Eq. \ref{e3}, we start the running of the couplings at the Z mass,
including the thresholds at $m_t$ and $\mu_{SUSY}$ (for the case
$\mu_{SUSY}<\mu_0$), and using the appropriate values of the coefficients
$b_i$ and $b_{ij}$'s. The modified minimal substraction scheme $(\overline{\mbox{MS}})\rightarrow\mbox{dimensional reduction }(\overline{\mbox{DR}})$ conversion factors
\begin{displaymath}
\Delta_i^{\mbox{conversion}}=-{{C_2\left(G_i\right)}\over{12\pi}}
\end{displaymath}
are included above $\mu_{SUSY}$. Beyond the compactification scale, $\mu_0$, the
effect of the extra dimensions on the running of the gauge couplings was
computed in \cite{dienes}. The couplings exibit an approximate power-law
evolution which, at the one-loop level, is given by \cite{dienes}
\begin{equation}
\label{e4}
\alpha_i^{-1}\left(\mu_0\right)= \alpha_i^{-1}\left(\Lambda\right)+
{{b_i-\tilde b_i}\over{\left(2\pi\right)}}\:\mbox{ln}{{\Lambda}\over{\mu_0}}
+{{\tilde b_i}\over{\left(2\pi\right)}}\:{{X_{\delta}}\over{\delta}}
\left[{\left({{\Lambda}\over{\mu_0}}\right)}^{\delta}-1\right]\: .
\end{equation}
The coefficients $\tilde b_i\equiv (\tilde b_1,\:\tilde b_2,\:\tilde b_3)$ are the appropriate $\beta$ function coefficients including
the contributions of the excited KK modes of all the particles living in the
$(4+\delta)$-dimensional space, and $\Lambda>\mu_0$. $\Lambda$ can be identified
with the GUT scale. $X_{\delta}$ is the volume of a $\delta$-dimensional unit sphere, given by
\begin{displaymath}
X_{\delta}={{2\pi^{\delta/2}}\over{\delta\:\Gamma\left(\delta/2\right)}},
\end{displaymath}
where $\Gamma$ is the Euler gamma function. In the running process we use
Eq. \ref{e3} and \ref{e4}, with the following input parameters:
\begin{eqnarray}
m_t&=&175\:\mbox{GeV},\nonumber\\ M_Z&=&91.187\:\mbox{GeV},\nonumber\\
\alpha_1^{-1}\left(M_Z\right)&=&58.9946,\nonumber\\
\alpha_2^{-1}\left(M_Z\right)&=&29.571.\nonumber
\end{eqnarray}
The value of $\alpha_3\left(M_Z\right)\equiv x$ was treated as a variable
to be solved for, along with $\Lambda/\mu_0\equiv y$ and $\alpha_{GUT}\equiv z$.
Thus, we have three equations for $\alpha_1\left(\mu\right)$, $\alpha_2\left(\mu\right)$, 
and $\alpha_3\left(\mu\right)$ [obtained by matching Eq. \ref{e3} and Eq. \ref{e4} at $\mu_0$], and three unknowns,
$x$, $y$, and $z$. These were solved for numerically, using the unification condition:
\begin{equation}
\alpha_1\left(\Lambda\right)=\alpha_2\left(\Lambda\right)=\alpha_3\left(\Lambda\right)=\alpha_{GUT}.
\end{equation}
For the case of $\mu_0<\mu_{SUSY}$, and also for the two-scale
compactification ($\mu_{10}$ and $\mu_{20}$), the evolution equations and
the $\beta$ function coefficients were adjusted appropriately. The values of
the coefficients are given for each case in Sec. \ref{sec3}, \ref{sec4}
and \ref{sec5}. The output of our calculations consists of
$\alpha_3\left(M_Z\right)$, $\Lambda$ and $\alpha_{GUT}$. This method has
the advantage that one can easily consider various possibilities for
$\mu_{SUSY}$ and $\mu_0$ (or $\mu_{10}$ and $\mu_{20}$). As a general
rule, the combinations that lead to the value of
$\alpha_3\left(M_Z\right)$ outside the $1\sigma$ range of the
experimental value ($0.1191\pm0.0018$) are discarded. So are combinations
that lead to the unified coupling outside the perturbative range
($\alpha_{GUT}\geq 1$).

Before the presentation of our results, we discuss the limitations
of this analysis. There are several threshold effects which we
have not included. These include supersymmetric thresholds and
higher-order logarithmic corrections, GUT thresholds, as well as
string thresholds from the oscillator states. The effect of these
thresholds on the gauge coupling evolution may not be
insignificant and could affect the unification in a non-negligible
way. For example, it is stated in \cite{dienes1} that a 6\%
threshold effect at the unification scale can give rise to precise
unification in the minimal scenario (our section \ref{sec3}).
Since we do not have a complete theory, and we do not know how to
calculate these threshold effects, we have implicitly assumed as
if the thresholds could be effectively included in the
experimental error bars. (Our conclusion holds if we increase the
error bars to $3\sigma$). However, it is quite possible that the
effect of these thresholds could be more significant.

\section{One com\-pac\-tification scale sce\-nar\-io with $\mu_{SUSY}\leq\mu_0$}
\label{sec3} As a first example we consider the minimal scenario of
Dienes, Dudas, and Gherghetta \cite{dienes}. We vary $\mu_{SUSY}$ from $1$ TeV up
to $2\times 10^3$ TeV and search for compactification scales
$\mu_0\geq\mu_{SUSY}$ that lead to acceptable predictions for
$\alpha_3\left(M_Z\right)$. Results are discarded if the prediction is off
by more than $1\sigma$. Our numerical results (see Table \ref{t1}) for the
case $\delta=1$, $\eta=0$ indicate that the lowest SUSY breaking scale for
which unification can occur is $\mu_{SUSY}=1.48\:$TeV, in which case the
compactification scale must be $\mu_0=3.27\times 10^{12}\:$TeV, leading to
unification at $\Lambda=6.25\times 10^{12}\:$TeV. Increasing
the number of extra dimensions has the effect of slightly increasing
these lower bounds on $\mu_{SUSY}$ and $\mu_0$.

\begin{figure}
\epsfbox[52 168 455 697]{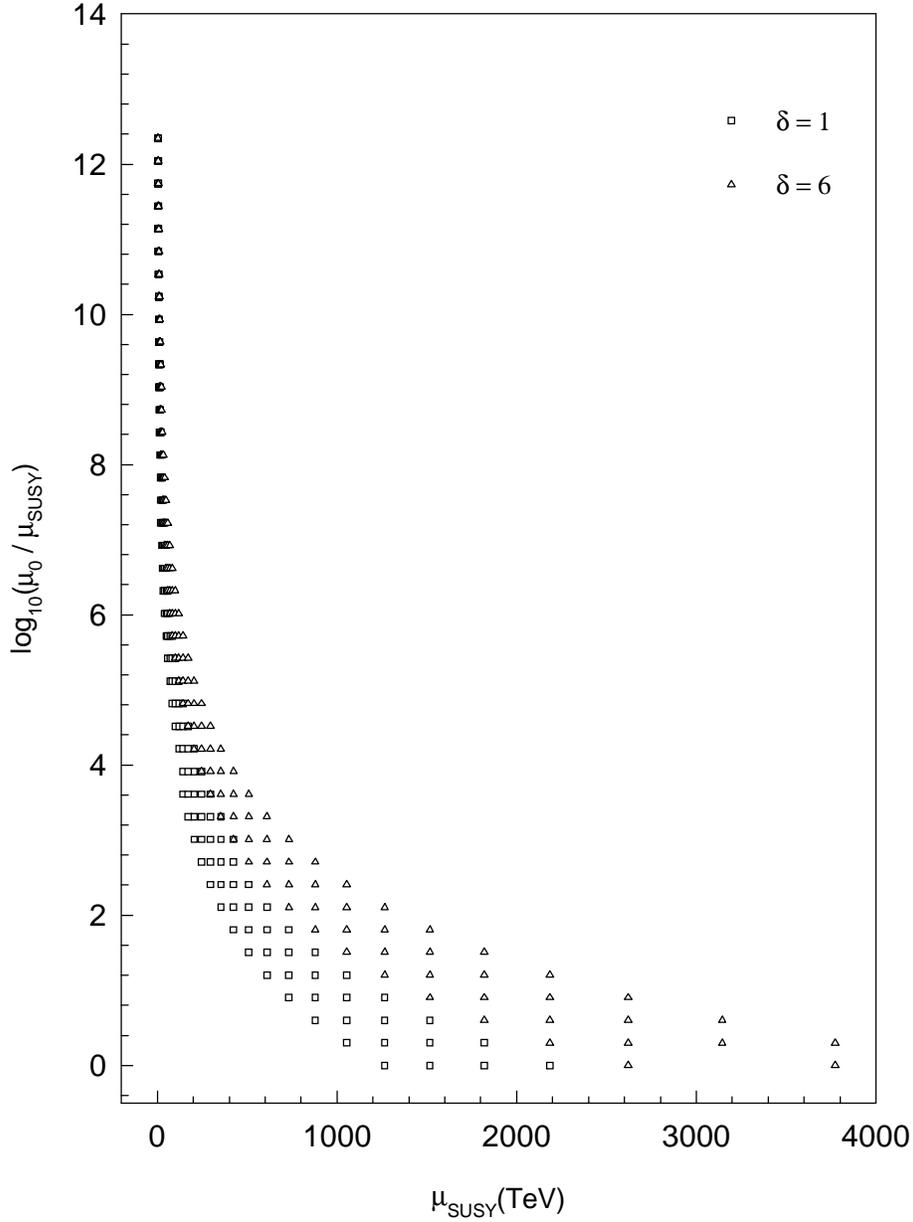} \caption{The ratio $\mu_0/\mu_{SUSY}$
plotted against various SUSY breaking scales, $\mu_{SUSY}$. Only results
within $1\sigma$ of  $\alpha_3\left(M_Z\right)$ are presented.
Unification is spoiled for points lying outside the corresponding bands.
}
\label{f1a}
\end{figure}

\begin{figure}
\epsfbox[66 92 459 622]{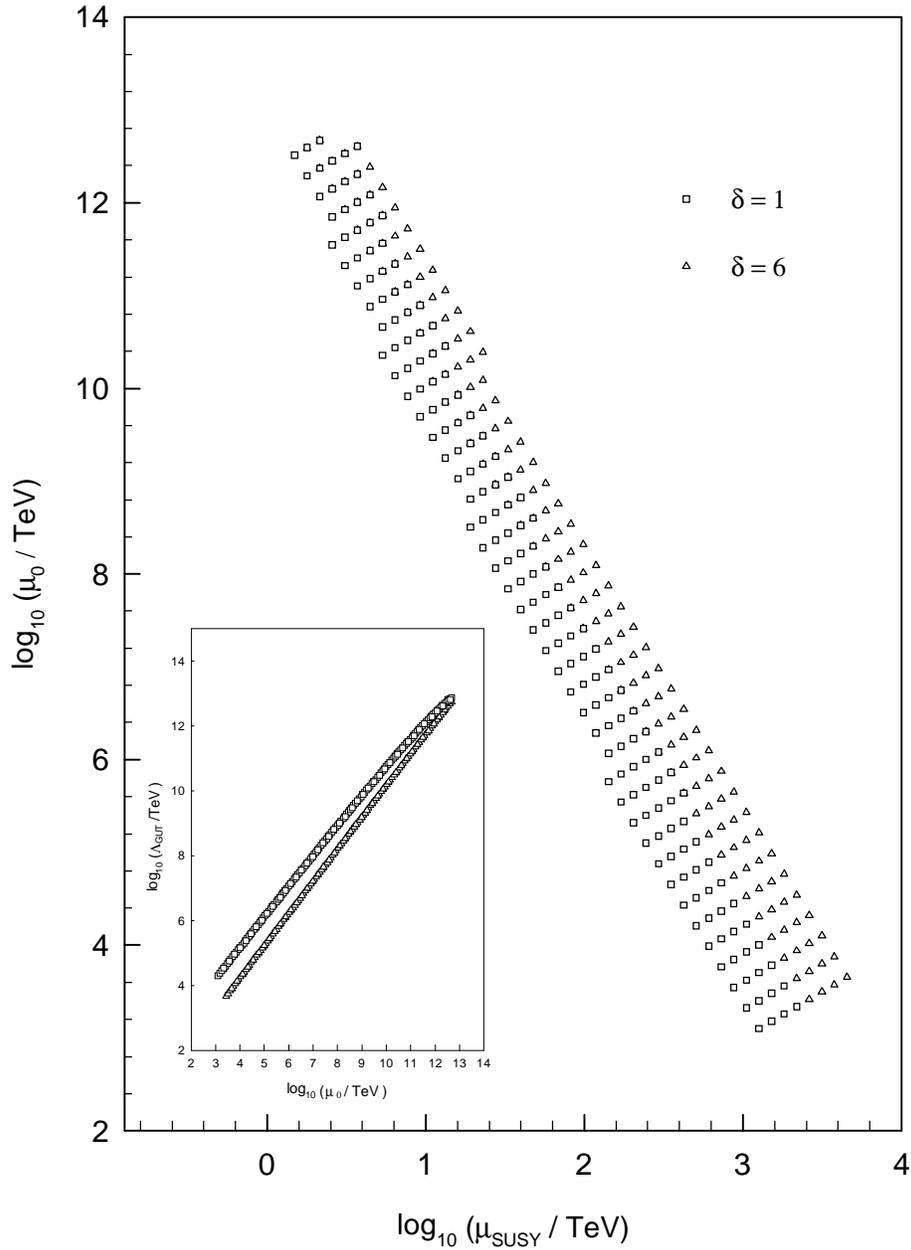} \caption{Scattered plot of the allowed
compactification scales, $\mu_0$, for various SUSY breaking scales,
$\mu_{SUSY}$. Only results within $1\sigma$ of $\alpha_3\left(M_Z\right)$
are presented. The same set of points as for the previous plot was used.
Unification is spoiled for points lying outside the corresponding bands.
The inset figure gives the unification scale against the
compactification scale.
}
\label{f1b}
\end{figure}

In Fig. \ref{f1a} we plot the ratio
$R=\mbox{log}_{10}\left({\mu_{SUSY}/\mu_0}\right)$ against $\mu_{SUSY}$.
The vertical and horizontal spreads in the figure represent the ranges for
which we get a solution at the $1\sigma$ range of
$\alpha_3\left(M_Z\right)$. As a general feature, as the SUSY breaking
scale increases, the compactification scale needed for unification
decreases, a ratio of approximately $1$ being obtained around
$\mu_{SUSY}\approx 1\times 10^3\:$TeV. This corresponds to the situation
in which supersymmetry is broken as soon as the extra dimensions
compactify. The same result is shown in Fig. \ref{f1b} where $\mu_0$ is
plotted against $\mu_{SUSY}$. The bands correspond to the regions in the
plane for which unification is achieved within $1\sigma$ range of
$\alpha_3\left(M_Z\right)$.
It is interesting to note that for the unification band $\mu_0$ is approximately
proportional to $\mu_{SUSY}^{-3}$.
For this case, our results are in agreement
with \cite{ross}. It can be concluded that there are no solutions leading
to both $\mu_{SUSY}$ and $\mu_0$ in the $100\:$TeV or less range. Allowing
$\eta\geq 1$ generations of matter fields to live in the
$(4+\delta)$-dimensional space drives the unified coupling $\alpha_{GUT}$
towards higher values while preserving unification (in agreement with
previous works).

\section{One compactification scale sce\-nar\-io with $\mu_{SUSY}\geq\mu_0$}
\label{sec4}
In this section we consider the posibility that the supersymmetry breaking
occurs at a scale higher than the compactification scale, $\mu_{SUSY}\geq\mu_0$.
For energies in the range $\mu_0\leq\mu\leq\mu_{SUSY}$ the theory is
nonsupersymmetric but the gauge and Higgs sectors of SM along with $\eta$
generations of matter fields exhibit KK excitations. The corresponding
contributions to the running are given by
\begin{displaymath}
\tilde b_i^{SM}=\left(1/10,-41/6,-21/2\right)+\eta\left(8/3,8/3,8/3\right).
\end{displaymath}
At $\mu_{SUSY}$ the theory becomes supersymmetric and additional KK
excitations of the sparticles lead to
\begin{displaymath}
\tilde b_i^{MSSM}=\left(3/5,-3,-6\right)+\eta\left(4,4,4\right).
\end{displaymath}
For the numerical analysis we choose various compactification scales
$\mu_0$ (starting in the TeV range) and search for SUSY breaking
scales that lead to acceptable predictions for $\alpha_3\left(M_Z\right)$
(within $1\sigma$ of the central experimental value).

\begin{figure}
\epsfbox[75 121 458 646]{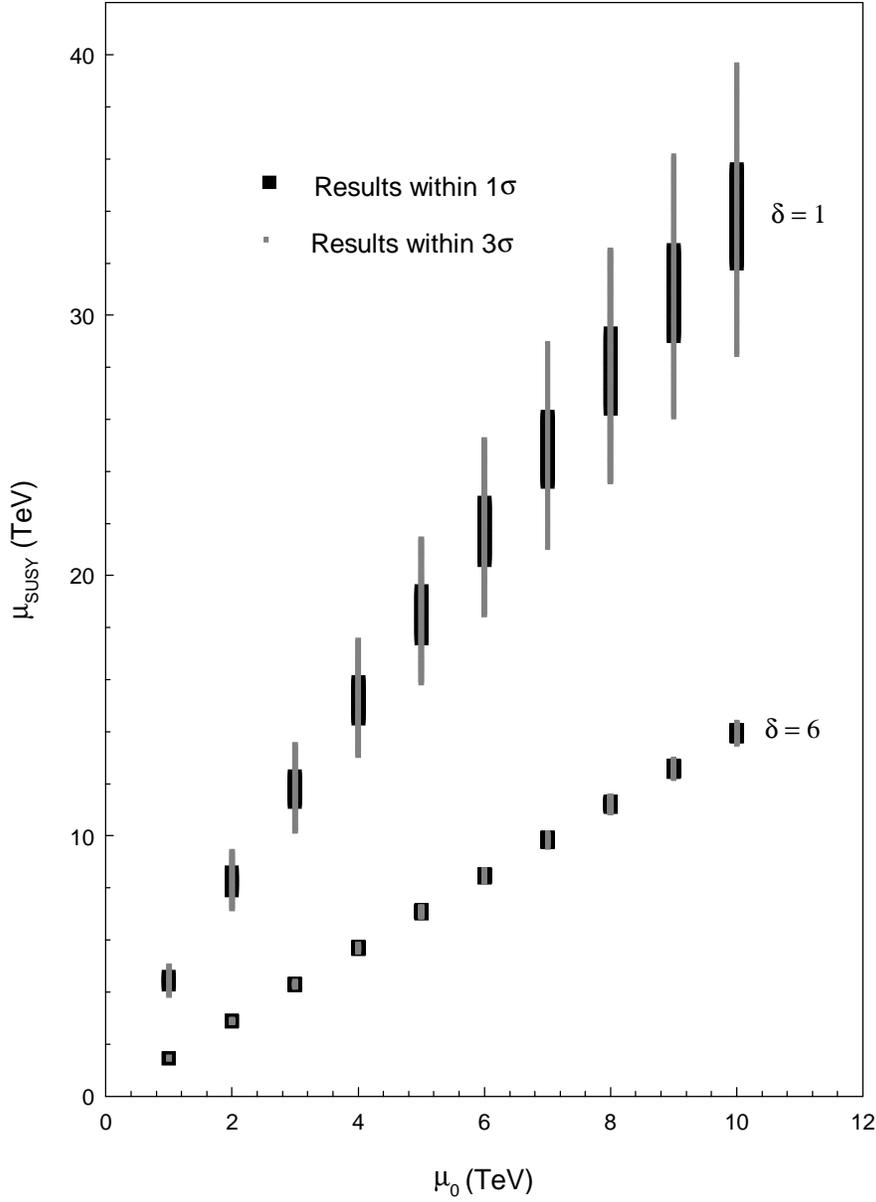} \caption{Allowed values of SUSY
breaking scale, $\mu_{SUSY}$, for various choices of $\mu_0$ in a scenario
with $\mu_0<\mu_{SUSY}$. Results within $1\sigma$ and $3\sigma$ of
$\alpha_3\left(M_Z\right)$ are presented, for $\delta=1$ and $\delta=6$.
Unification is spoiled if $\mu_{SUSY}$ lies outside the corresponding
vertical spreads shown in the plot.}
\label{f2}
\end{figure}

For the simplest case, $\eta=0$, the results
are shown in Fig. \ref{f2} where the allowed values of $\mu_{SUSY}$ are
plotted against the corresponding compactification scale $\mu_0$, for
$\delta=1$ and $\delta=6$. Relevant
numerical results are presented in Table \ref{t2}.

As a generic feature, each compactification
scale corresponds to a specific range of $\mu_{SUSY}$ that are needed for
unification and are consistent with low-energy experimental data.
The length of these intervals is, of course, determined by our requirement
of $1\sigma$ (or $3\sigma$) agreement with the experimental value of $\alpha_3\left(M_Z\right)$
but it is found to increase with $\mu_0$.
The fact that the upper bound of these ranges is finite shows that,
within this model,
supersymmetry is in fact needed for unification. Unification cannot
occur within the SM spectrum. This was also noticed in \cite{dienes} for the case
$\mu_{SUSY}<\mu_0$.

This scenario is particularly appealing from the experimental point of
view. Ignoring possible constraints on $\mu_0$ we consider a
compactification scale as low as $\mu_0=1\:$TeV which enforces
$\mu_{SUSY}=4.5\:$TeV and $\mu_{SUSY}=1.46\:$TeV for $\delta=1$ and
$\delta=6$, respectively. This leads to unification at $\Lambda=75.2\:$TeV for
$\delta=1$ and $\Lambda=2.68\:$TeV for $\delta=6$.
A more realistic case would be $\mu_0=3\:$TeV, $\mu_{SUSY}=11.9\:$TeV
with unification at $\Lambda=198\:$TeV for $\delta=1$ or $\mu_{SUSY}=4.3\:$TeV
with unification at $\Lambda=7.86\:$TeV for $\delta=6$. Needless to say, these cases are well within the LHC
reach and can be investigated at future experiments. The case $\eta=3$,
not present in Table \ref{t2}, led to negative unified coupling for the range
of $\mu_0$ shown in Fig. \ref{f2}.

Let us point out one caveat in the scenario presented. We have
included only the one-loop contribution to the evolution of the
couplings in the energy scale $\mu>\mu_0$. The corrections to the
power-law evolution of the couplings due to the contribution of 
higher loops are small if the theory is supersymmetric, as shown
in \cite{kaku}. Thus in our scenario these corrections are small
in the region $\mu_{SUSY}<\mu<\Lambda$. However, it is possible
that such corrections could significantly affect our results in
the region $\mu_0<\mu<\mu_{SUSY}$ due to the absence of
supersymmetry in this region. This is especially true because
realistically, the SUSY particles are not expected to be
degenerate in mass at $\mu_{SUSY}$. Since in the region
$\mu_0<\mu<\mu_{SUSY}$ the running of the couplings is
approximately power law, the effect of the separate thresholds on
the running, in principle, may be greater than in the logarithmic
case. However, in the absence of a complete theory, it is not
possible to calculate such effects. We note that this problem does
not appear for $\mu_0\approx\mu_{SUSY}\approx 10^3$ TeV for which
we do get unification of the gauge couplings.

We conclude this section with a few remarks. It was suggested in
the literature \cite{dienes} that the compactification scale could be
identified with the SUSY breaking scale. Our results in this section and
Sec. \ref{sec3}
indicate that, if this is the case, then this
common scale cannot be lower than $10^6\:$GeV (around $10^6\:$Gev the ratio
$\mu_{SUSY}/\mu_0$ required for unification approaches $1$ in both scenarios).

One question needs to be addressed here. Does string theory allow a
scenario in which the compactification scale is lower than the
SUSY breaking scale? In this case, SUSY has to be broken in a 
higher dimension before compactification. There are several
possibilities for that to happen. One possibility is a string
solution in which SUSY is broken at the string level. In general,
non-SUSY string solutions are unstable. String theory prefers
vacua which are supersymmetric. Dilaton and other modulii tend to
run away to infinity, and restore SUSY. However, given the reach
complexities and possibilities in string theory, such a scenario
can not be ruled out. A second possibility is the gaugino
condensation in higher dimensional gauge theory. The gauge
coupling could be of the order of unity, causing gaugino condensation and
breaking $N=2$ (or even $N=1$) SUSY, before compactification to
four dimensions. Yet another possibility is that the SM particles
(plus their SUSY partners) live in a non-Bogomol'nyi-Prasad-Sommerfield (BPS) brane which is stable
but does not preserve supersymmetry at all \cite{last}. Thus, we
conclude that a scenario with $\mu_{SUSY}>\mu_0$ is not totally
crazy, although such a scenario is probably not as well motivated
as the $\mu_{SUSY}<\mu_0$ case.
\section{Two scale compactification scenarios}
\label{sec5}
In the analysis of Sec. \ref{sec3} and \ref{sec4} we assumed that the compactification
of the extra dimensions takes place at a single mass scale, $\mu_0$.
However, the possibility exists that the different extra dimensions compactify at
different mass scales. Also, particles with different gauge quantum numbers may belong
to different D branes associated with different compactification scales. This section is devoted
to numerical analyses of such scenarios with two different mass scales,
$\mu_{10}$ and $\mu_{20}$ with $\mu_{10}<\mu_{20}$. In these models the MSSM
spectrum (or only a subset of it) is split up into two parts, with the first
part developing KK excitations at the first compactification scale $\mu_{10}$
and with the remainder contributing only after the second scale $\mu_{20}$ is crossed.

In all the subsequent cases the SUSY breaking scale is assumed to be lower than
$\mu_{10}$.
For practical
purposes we restrict ourselves to compactification scales $\mu_{10}$
that are within the LHC reach and to $\delta=1$ extra dimensions. Only results
that lead to predictions of $\alpha_3(M_Z)$ within $1\sigma$
of the central experimental value are presented.

In what follows we consider several scenarios in which the splitting of the
MSSM gauge sector is based on color. Relevant numerical results for these models are
presented in Table \ref{t3} and the $\beta$-function coefficients corresponding to
the two compactification scales for the cases A, B, C, D presented below, are given by
\begin{eqnarray}
\tilde b_i^{\left(10\right)}&=&\left(0,0,-6\right),\label{bet}\\
\tilde b_i^{\left(20\right)}&=&\left(0,-4,-6\right)+\eta\left(4,4,4,\right)\nonumber
\end{eqnarray}
with the appropriate choice of $\eta$.\\[.9cm]
\bf Case A\hfill\\[\baselineskip]\rm
$\mu_{10}\rightarrow$ SU(3)\hfill\newline
$\mu_{20}\rightarrow$ SU(3)$\otimes$SU(2)$\otimes$U(1)\hfill\\[\baselineskip]
The notation is that only the gluons (along with their SUSY partners) develop
KK excitations at $\mu_{10}$, while the full MSSM gauge sector contribute above $\mu_{20}$.
The $\beta$-function coefficients are given by Eq. \ref{bet} with $\eta=0$.
For SUSY breaking scales in the TeV range and $\mu_{10}$ within the
reach of LHC ($\leq 14\:$TeV), a ratio $\mu_{20}/\mu_{10}$ of about 7
is needed in order to achieve unification [with the prediction for $\alpha_3(M_Z)$ within
$1\sigma$ of the central experimental value].
The unification scale is as low as
$4\times 10^2\:$TeV.
Note that for this scenario the value of the couplings at the unification
scale ($\alpha_{GUT}\approx 0.015$) is significantly smaller than
$\alpha_3\left(M_Z\right)$ and well within the perturbative regime. As a general feature, attempts to bring
the compactification scale $\mu_{10}$ down to $\mu_{SUSY}$ (at fixed
$\mu_{20}/\mu_{10}$) tend to drive the unified coupling towards higher values.\\[\baselineskip]
\bf Case B\hfill\\[\baselineskip]\rm
$\mu_{10}\rightarrow$ SU(3),\hfill\\
$\mu_{20}\rightarrow$ SU(3)$\otimes$SU(2)$\otimes$U(1)$\;\oplus\; 1\:\mbox{generation of matter fields}$ \hfill\\[\baselineskip]
[$\eta =1$ in Eq. \ref{bet}]. The addition of $\eta=1$ generation of matter fields at $\mu_{10}$ preserves unification
while increasing the coupling at the unification scale ($\alpha_{GUT}\approx 0.032$).
This case shares all the features of the previous one.\\[\baselineskip]
\bf Case C\hfill\\[\baselineskip]\rm
$\mu_{10}\rightarrow$ SU(3),\hfill\\
$\mu_{20}\rightarrow$ SU(3)$\otimes$SU(2)$\otimes$U(1)$\;\oplus\; 2\:\mbox{generations of matter fields}$\hfill\\[\baselineskip]
[$\eta =2$ in Eq. \ref{bet}]. With an MSSM spectrum at the TeV scale we found
that this scenario does not lead to unification for $\mu_{10}$ within the LHC reach
[although a \it mathematical \rm unification is achieved, either
the unified coupling $\alpha_{GUT}$ has unphysical values or the prediction for
$\alpha_3(M_Z)$ is outside $3\sigma$ of the experimental value]. However, extending
the range of $\mu_{10}$ beyond the reach of LHC we found that unification can be achieved for
$\mu_{10}\geq 5\times 10^2\:\mbox{TeV}$ and only for $\mu_{20}/\mu_{10}\approx 5.5$.
The unification scale can be as low as $\Lambda = 7.8\times 10^4\:\mbox{TeV}$ and the unified coupling is
in the perturbative regime.\\[\baselineskip]
\bf Case D\hfill\\[\baselineskip]\rm
$\mu_{10}\rightarrow$ SU(3),\hfill\\
$\mu_{20}\rightarrow$ SU(3)$\otimes$SU(2)$\otimes$U(1)$\;\oplus\; 3\:\mbox{generations of matter fields}$\hfill\\[\baselineskip]
[$\eta =3$ in Eq. \ref{bet}]. This case is similar to Case C. A minimum compactification scale
of $\mu_{10}\approx 7\times 10^7\:\mbox{TeV}$ and a ratio $\mu_{10}/\mu_{20}\approx 3.4$ are
required for unification. Consequently, the unification scale is pushed towards about
$\Lambda=3.1\times 10^9\:\mbox{TeV}$.

In Table \ref{cases} we list several other cases that were investigated
but found {\it not} to give results of interest for future experiments at LHC.

Several conclusions can be drawn from the results above. Most importantly,
the 2-scale scenarios allow for very low compactification scales (in the
TeV range) even for the case in which the SUSY braking scale is lower
than the compactification scale. This was not possible in one-scale
scenarios. Moreover, results with $\mu_{SUSY}=\mu_{10}\:\approx$ few TeV are
obtained, which encourages the identification of the SUSY breaking scale with
the compactification scale. Specification of $\mu_{SUSY}$ along with the
requirement that the first threshold is within the LHC reach, completely
determined the second threshold as well as the unification scale [of
course, with small variations determined by the error bar on the experimental
value of $\alpha_3\left(M_Z\right)$].

\section{Conclusions}
\label{sec6} In this work we have made a detailed investigation for the
unification of the gauge couplings in MSSM with extra dimensions. We do
not extend the gauge group or the field content (except for those required
by the higher dimensions). In the previous studies, it was implicitly
assumed that the SUSY breaks at four dimensions before the
decompactification, and thus the scale of SUSY breaking, $\mu_{SUSY}$ is
lower than the decompactification scale, $\mu_0$. In this case, it was
observed that the three gauge couplings do not unify [satisfying the
experimental range of $\alpha_3\left(M_Z\right)$] with both $\mu_{SUSY}$
and $\mu_0$ less than a few tens of a TeV. We have investigated several new
scenarios for which the couplings unify with both $\mu_{SUSY}$ and $\mu_0$
in the few TeV scale. One particularly interesting scenario is when SUSY
is broken at higher dimension [either through string dynamics or via
gaugino condensation or in a non-Bogomol'nyi-Prasad-Sommerfield (BPS) brane] before decompactification, so that
$\mu_{SUSY}>\mu_0$. In this case we obtained gauge coupling unification
with both $\mu_{SUSY}$ and $\mu_0$ in the few TeV scale. This is very
exciting, since for this scenario, LHC ($\sqrt{s}=14\:$TeV) will be able
to probe experimentally the existence of these compact dimensions. The
direct experimental test will be the observation of the low-lying KK
resonance of SM particles, or the off-shell effect of these particles via
the usual SM processes. A family of two scale compactification scenarios in which the MSSM gauge sector
is split into its colored and uncolored subsets was also considered. It was found
that with $\eta=0,1$ matter generations contributing above the second scale
$\mu_{20}$, the unification can be achieved with both $\mu_{SUSY}$ and
$\mu_{10}$ in the few TeV scale. In all cases unification can be achieved
only for a specific narrow range of the ratio
$\mu_{20}/\mu_{10}$.
\\[2cm]
\Large\bf Acknowledgments
\rm\normalsize\\[.5cm] We wish to thank K. S. Babu, J. Lykken and G. Gabadadze for
useful discussions. This work was supported in part by the U.S. Department
of Energy, Grant Number DE-FG03-98ER41076.

\bibliographystyle{unsrt}

\newpage

\begin{table}
\begin{center}
\begin{tabular}{|l|l|l|l|l|l|l|l|}
\hline
$\delta$&$\eta$&$\mu_{SUSY}$&$\mu_0$&$\alpha_3\left(M_Z\right)$&$\Lambda/\mu_0$&$\Lambda$&$\alpha_{GUT}$\\
\hline\hline
$1 $&$0 $&$1.48\times 10^3 $&$3.27\times 10^{15} $&$0.1208$&$1.91 $&$6.25\times 10^{15} $&$0.0384 $\\
\hline
$1 $&$0 $&$5.32\times10^3 $&$2.29\times 10^{13} $&$0.1208 $&$5.14 $&$1.18\times 10^{14} $&$0.0330$\\
\hline\hline
$6 $&$0 $&$1.78\times 10^3 $&$3.92\times 10^{15} $&$0.1206$&$1.20 $&$4.71\times 10^{15} $&$0.0379$\\
\hline
$6 $&$0 $&$5.32\times 10^3$&$3.66\times 10^{14} $&$0.1193 $&$1.37 $&$5.00\times 10^{14} $&$0.0345 $\\
\hline\hline
$1 $&$1 $&$5.32\times 10^3 $&$2.29\times 10^{13} $&$0.1208$&$5.14 $&$1.18\times 10^{14} $&$0.0383 $\\
\hline
$1 $&$2 $&$5.32\times 10^3 $&$2.29\times 10^{13} $&$0.1208$&$5.14 $&$1.18\times 10^{14} $&$0.0457 $\\
\hline
$1$&$3 $&$5.32\times 10^3 $&$2.29\times 10^{13} $&$0.1208 $&$5.14$&$1.18\times 10^{14} $&$0.0567 $\\
\hline
\end{tabular}
\end{center}
\caption{A few relevant numerical results for a one threshold scenario
with $\mu_{SUSY}<\mu_0$. The behavior under changes of $\delta$ and $\eta$
is shown. All the mass scales are in GeV units. Relevant plots are presented in
Fig. \protect\ref{f1a} and \protect\ref{f1b}.}
\label{t1}
\end{table}

\newpage

\begin{table}
\begin{center}
\begin{tabular}{|l|l|l|l|l|l|l|l|l|}
\hline
$\delta$&$\eta$&$\mu_0$&$\mu_{SUSY}/\mu_0$&$\mu_{SUSY}$&$\alpha_3\left(M_Z\right)$&$\Lambda/\mu_0$&$\Lambda$&$\alpha_{GUT}$\\
\hline\hline
$1 $&$0 $&$1\times 10^3 $&$4.5 $&$4.5\times 10^3 $&$0.1187 $&$75.2 $&$7.52\times 10^4 $&$0.0197$\\
\hline
$1 $&$0 $&$2\times 10^3 $&$4.2 $&$8.3\times 10^3 $&$0.1190 $&$69.3 $&$1.39\times 10^5 $&$0.0199 $\\
\hline
$1 $&$0 $&$3\times 10^3 $&$4.0 $&$1.19\times 10^4 $&$0.1190 $&$66.0 $&$1.98\times 10^5 $&$0.0200 $\\
\hline
$1 $&$0 $&$4\times 10^3 $&$3.8 $&$1.53\times 10^4 $&$0.1191 $&$63.6 $&$2.55\times 10^5 $&$0.0200 $\\
\hline
$1 $&$0 $&$5\times 10^3 $&$3.8 $&$1.86\times 10^4 $&$0.1191 $&$61.8 $&$3.09\times 10^5 $&$0.0201 $\\
\hline
$1 $&$0 $&$7\times 10^3 $&$3.6 $&$2.50\times 10^4 $&$0.1191 $&$59.2 $&$4.14\times 10^5 $&$0.0201 $\\
\hline
$1 $&$0 $&$9\times 10^3 $&$3.5 $&$3.11\times 10^4 $&$0.1191 $&$57.1 $&$5.14\times 10^5 $&$0.0202 $\\
\hline
\hline
$6 $&$0 $&$1\times 10^3 $&$1.46 $&$1.46\times 10^3 $&$0.1201 $&$2.68 $&$2.68\times 10^3 $&$0.0187 $\\
\hline
$6 $&$0 $&$2\times 10^3 $&$1.45 $&$2.89\times 10^3 $&$0.1194 $&$2.64 $&$5.29\times 10^3 $&$0.0188 $\\
\hline
$6 $&$0 $&$3\times 10^3 $&$1.43 $&$4.3\times 10^3 $&$0.1194 $&$2.62 $&$7.86\times 10^3 $&$0.0189 $\\
\hline
$6 $&$0 $&$4\times 10^3 $&$1.43 $&$5.71\times 10^3 $&$0.1190 $&$2.61 $&$1.04\times 10^4 $&$0.0190 $\\
\hline
$6 $&$0 $&$5\times 10^3 $&$1.42 $&$7.10\times 10^3 $&$0.1191 $&$2.59 $&$1.29\times 10^4 $&$0.0190 $\\
\hline
$6 $&$0 $&$7\times 10^3 $&$1.41 $&$9.86\times 10^3 $&$0.1192 $&$2.57 $&$1.80\times 10^4 $&$0.0191 $\\
\hline
$6 $&$0 $&$9\times 10^3 $&$1.40 $&$1.26\times 10^4 $&$0.1191 $&$2.56 $&$2.3\times 10^4 $&$0.0192 $\\
\hline\hline
$1 $&$1 $&$1\times 10^3 $&$4.5 $&$4.5\times 10^3 $&$0.1187 $&$75.2 $&$7.52\times 10^4 $&$0.0332 $\\
\hline
$1 $&$2 $&$1\times 10^3 $&$4.5 $&$4.5\times 10^3 $&$0.1187 $&$75.2 $&$7.52\times 10^4 $&$0.1040$\\
\hline
$6 $&$1 $&$1\times 10^3 $&$1.46 $&$1.46\times 10^3 $&$0.1201 $&$2.68 $&$2.68\times 10^3 $&$0.0327 $\\
\hline
$6 $&$2 $&$1\times 10^3 $&$1.46 $&$1.46\times 10^3 $&$0.1201 $&$2.68 $&$2.68\times 10^3 $&$0.1300 $\\
\hline
\end{tabular}
\end{center}
\caption{Numerical results for the 1-scale scenario with
$\mu_{SUSY}\geq\mu_0$. The compactification scales $\mu_0$ were
taken as input and the allowed values of SUSY breaking scales
were determined numerically. The behaviour under changes
of $\delta$ and $\eta$ is shown. All the mass scales are in GeV.
See Fig. \protect\ref{f2} for a relevant plot.}
\label{t2}
\end{table}
\newpage

\begin{table}
\begin{center}
\begin{tabular}{|l|l|}
\hline
$\mu_{10}$&SU(3)\\
\hline
$\mu_{20}$&SU(3)$\otimes$SU(2)$\otimes$U(1) $\oplus$ H\\
\hline\hline
$\mu_{10}$&SU(3)\\
\hline
$\mu_{20}$&SU(3)$\otimes$SU(2)$\otimes$U(1) $\oplus$ 3(L, E)\\
\hline\hline
$\mu_{10}$&SU(3)\\
\hline
$\mu_{20}$&SU(3)$\otimes$SU(2)$\otimes$U(1) $\oplus$ 3(L, Q)\\
\hline\hline
$\mu_{10}$&SU(3) $\otimes$ U(1) $\oplus$ 3(U, D)\\
\hline
$\mu_{20}$&SU(3)$\otimes$SU(2)$\otimes$U(1) $\oplus$ 3(Q, U, D, L, E)\\
\hline\hline
$\mu_{10}$&SU(3) $\otimes$ U(1) $\oplus$ 3(U, D)\\
\hline
$\mu_{20}$&SU(3)$\otimes$SU(2)$\otimes$U(1) $\oplus$ 3(Q, U, D, L, E) $\oplus$ H\\
\hline\hline
$\mu_{10}$&SU(2) \\
\hline
$\mu_{20}$&SU(3)$\otimes$SU(2)$\otimes$U(1) \\
\hline\hline
$\mu_{10}$&SU(2) $\otimes$ U(1) $\oplus$ 3(L, E) $\oplus$ H\\
\hline
$\mu_{20}$&SU(3)$\otimes$SU(2)$\otimes$U(1) $\oplus$ 3(Q, U, D, L, E) $\oplus$ H\\
\hline
\end{tabular}
\end{center}
\caption{Two compactification scale scenarios which DO NOT lead to unification with
both $\mu_{SUSY}$ and $\mu_{10}$ within the reach of LHC. }
\label{cases}
\end{table}
\newpage

\begin{table}
\begin{center}
\begin{tabular}{|l|l|l|l|l|l|l|l|l|}
\hline
$\eta$&$\mu_{SUSY}$&$\mu_{10}$&$R$&$\mu_{20}$&$\alpha_3\left(M_Z\right)$&$\Lambda/\mu_{10}$&$\Lambda$&$\alpha_{GUT}$\\
\hline\hline
0&$1\times 10^3 $&$2\times 10^3 $&$7.2 $&$1.44\times 10^4 $&$0.1189 $&$212 $&$4.23\times 10^5 $&$0.0156 $\\
\hline
0&$2\times 10^3 $&$2\times 10^3 $&$7.2 $&$1.44\times 10^4 $&$0.1176 $&$211 $&$4.21\times 10^5 $&$0.0155 $\\
\hline
0&$2\times 10^3 $&$4\times 10^3 $&$7.2 $&$2.88\times 10^4 $&$0.1196 $&$207 $&$8.28\times 10^5 $&$0.0157 $\\
\hline
0&$2\times 10^3 $&$6\times 10^3 $&$7.2 $&$4.32\times 10^4 $&$0.1209 $&$205 $&$1.23\times 10^6 $&$0.0158 $\\
\hline
0&$3\times 10^3 $&$3\times 10^3 $&$7.2 $&$2.16\times 10^4 $&$0.1179 $&$208 $&$6.23\times 10^5 $&$0.0156 $\\
\hline
0&$3\times 10^3 $&$5\times 10^3 $&$7.2 $&$3.60\times 10^4 $&$0.1195 $&$205 $&$1.03\times 10^6 $&$0.0157 $\\
\hline
0&$3\times 10^3 $&$7\times 10^3 $&$7.2 $&$5.04\times 10^4 $&$0.1205 $&$204 $&$1.43\times 10^6 $&$0.0158 $\\
\hline
0&$5\times 10^3 $&$5\times 10^3 $&$7.2 $&$3.60\times 10^4 $&$0.1184 $&$205 $&$1.02\times 10^6 $&$0.0157 $\\
\hline
0&$5\times 10^3 $&$7\times 10^3 $&$7.2 $&$5.04\times 10^4 $&$0.1194 $&$203 $&$1.42\times 10^6 $&$0.0158 $\\
\hline
0&$5\times 10^3 $&$9\times 10^3 $&$7.2 $&$6.48\times 10^4 $&$0.1202 $&$202 $&$1.81\times 10^6 $&$0.0158 $\\
\hline
\hline
1&$1\times 10^3 $&$2\times 10^3 $&$7.2 $&$1.44\times 10^4 $&$0.1189 $&$212 $&$4.23\times 10^5 $&$0.0332 $\\
\hline
1&$2\times 10^3 $&$2\times 10^3 $&$7.2 $&$1.44\times 10^4 $&$0.1176 $&$211 $&$4.21\times 10^5 $&$0.0327 $\\
\hline
1&$2\times 10^3 $&$4\times 10^3 $&$7.2 $&$2.88\times 10^4 $&$0.1196 $&$207 $&$8.28\times 10^5 $&$0.0329 $\\
\hline
1&$2\times 10^3 $&$6\times 10^3 $&$7.2 $&$4.32\times 10^4 $&$0.1209 $&$205 $&$1.23\times 10^6 $&$0.0329 $\\
\hline
1&$3\times 10^3 $&$3\times 10^3 $&$7.2 $&$2.16\times 10^4 $&$0.1179 $&$208 $&$6.23\times 10^5 $&$0.0325 $\\
\hline
1&$3\times 10^3 $&$5\times 10^3 $&$7.2 $&$3.60\times 10^4 $&$0.1195 $&$205 $&$1.03\times 10^6 $&$0.0326 $\\
\hline
1&$3\times 10^3 $&$7\times 10^3 $&$7.2 $&$5.04\times 10^4 $&$0.1205 $&$204 $&$1.43\times 10^6 $&$0.0327 $\\
\hline
1&$5\times 10^3 $&$5\times 10^3 $&$7.2 $&$3.60\times 10^4 $&$0.1184 $&$205 $&$1.02\times 10^6 $&$0.0322 $\\
\hline
1&$5\times 10^3 $&$7\times 10^3 $&$7.2 $&$5.04\times 10^4 $&$0.1194 $&$203 $&$1.42\times 10^6 $&$0.0323 $\\
\hline
1&$5\times 10^3 $&$9\times 10^3 $&$7.2 $&$6.48\times 10^4 $&$0.1202 $&$202 $&$1.81\times 10^6 $&$0.0324 $\\
\hline\hline
2&$3\times 10^3 $&$3.1\times 10^7 $&$5.1 $&$1.55\times 10^8 $&$0.1174 $&$99 $&$3.04\times 10^9 $&$0.1364 $\\
\hline
2&$3\times 10^3 $&$5.3\times 10^7 $&$5.1 $&$2.69\times 10^8 $&$0.1190 $&$98 $&$5.15\times 10^9 $&$0.1302 $\\
\hline
2&$3\times 10^3 $&$9.1\times 10^7 $&$5.1 $&$4.64\times 10^8 $&$0.1206 $&$96 $&$8.73\times 10^9 $&$0.1244 $\\
\hline
2&$3\times 10^3 $&$5.5\times 10^5 $&$6 $&$3.31\times 10^6 $&$0.1175 $&$142 $&$7.81\times 10^7 $&$0.3285 $\\
\hline
2&$3\times 10^3 $&$9.5\times 10^5 $&$6 $&$5.72\times 10^6 $&$0.1191 $&$139 $&$1.33\times 10^8 $&$0.2935 $\\
\hline
2&$3\times 10^3 $&$1.7\times 10^6 $&$6 $&$9.89\times 10^6 $&$0.1208 $&$137 $&$2.26\times 10^8 $&$0.2652 $\\
\hline\hline
3&$3\times 10^3 $&$4.8\times 10^{11} $&$3 $&$1.43\times 10^{12} $&$0.1177 $&$30 $&$1.4\times 10^{13} $&$0.1386 $\\
\hline
3&$3\times 10^3 $&$8.2\times 10^{11} $&$3 $&$2.27\times 10^{12} $&$0.1193 $&$29 $&$2.4\times 10^{13} $&$0.1248 $\\
\hline
3&$3\times 10^3 $&$1.4\times 10^{12} $&$3 $&$4.27\times 10^{12} $&$0.1209 $&$28 $&$3.9\times 10^{13} $&$0.1135 $\\
\hline
3&$3\times 10^3 $&$7.7\times 10^{10} $&$3.4 $&$2.62\times 10^{11} $&$0.1177 $&$40 $&$3.1\times 10^{12} $&$0.3127 $\\
\hline
3&$3\times 10^3 $&$9.2\times 10^{10} $&$3.4 $&$3.14\times 10^{11} $&$0.1183 $&$40 $&$3.7\times 10^{12} $&$0.2884 $\\
\hline
3&$3\times 10^3 $&$1.3\times 10^{11} $&$3.4 $&$4.53\times 10^{11} $&$0.1193 $&$39 $&$5.2\times 10^{12} $&$0.2497 $\\
\hline
\end{tabular}
\end{center}
\caption{Numerical results for two scale compactification scenarios.
The cases $\eta=0,1,2,3$ correspond to cases A,B,C,D respectively and $R=\mu_{20}/\mu_{10}$.
The number of extra dimensions is $\delta=1$
and the mass scales are in GeV.}
\label{t3}
\end{table}

\clearpage

\clearpage

\end{document}